\documentclass[12pt]{article}
\usepackage[a4paper, total={7in, 10in}]{geometry}
\usepackage[parfill]{parskip}
\usepackage{physics, tensor, float, subcaption}
\usepackage{graphicx}
\graphicspath{ {Plots/} }
\usepackage{jhep-mod}
\usepackage{bm}
\usepackage{soul}
\usepackage{amssymb,amsmath,amsthm}
\usepackage{mathrsfs}
\usepackage[utf8]{inputenc}
\usepackage{enumerate}
\usepackage{bigints}
\usepackage{xcolor}
\usepackage{appendix}
\usepackage{graphicx}
\usepackage{float}
\usepackage{tikz}
\usepackage{setspace}
\usepackage{cancel}
\usepackage{tocbasic}
\DeclareTOCStyleEntry[
  beforeskip=.2em plus 1pt,% default is 1em plus 1pt
  pagenumberformat=\textbf
]{tocline}{section}
\definecolor{purple}{rgb}{1,0,1}
\definecolor{lime}{HTML}{A6CE39} % needs xcolor
\definecolor{lime}{HTML}{A6CE39}
\newcommand{\orcidicon}{%
	\begin{tikzpicture}
	\draw[lime, fill=lime] (0,0) 
		circle [radius=0.16] 
		node[white] {{\fontfamily{qag}\selectfont \tiny ID}};
	\draw[white, fill=white] (-0.0625,0.095) 
		circle [radius=0.007];
	\end{tikzpicture}
	\hspace{-5mm}
}
\newcommand\orcidJoshua{{\href{https://orcid.org/0000-0003-1200-7261}{\orcidicon}}}
\newcommand\orcidThomas{{\href{https://orcid.org/0000-0002-0314-4136}{\orcidicon}}}
\newcommand\orcidAlex{{\href{https://orcid.org/0000-0002-1763-3563}{\orcidicon}}}
\newcommand\orcidMatt{{\href{https://orcid.org/0000-0003-1088-6485}{\orcidicon}}}
\begin{document}
%========================================================

%========================================================
%========================================================
\title{\vspace{-45pt}\huge{
{Darboux diagonalization of the spatial 3-metric in Kerr spacetime}
}}
%========================================================
%========================================================
%========================================================
\author{
\Large
Joshua Baines\!\orcidJoshua\!, Thomas Berry\!\orcidThomas\!, Alex Simpson\!\orcidAlex\!,
\\{\sf  and} Matt Visser\!\orcidMatt}
%========================================================
%========================================================
%========================================================
%========================================================
\affiliation{School of Mathematics and Statistics, Victoria University of Wellington, 
\\
\null\qquad PO Box 600, Wellington 6140, New Zealand.}
%========================================================
%========================================================
\emailAdd{joshua.baines@sms.vuw.ac.nz}
\emailAdd{thomas.berry@sms.vuw.ac.nz}
\emailAdd{alex.simpson@sms.vuw.ac.nz}
\emailAdd{matt.visser@sms.vuw.ac.nz}
%========================================================
%========================================================

\abstract{
\vspace{1em}

The astrophysical importance of the Kerr spacetime cannot be overstated. 
Of the currently known exact solutions to the Einstein field equations, the Kerr spacetime stands out in terms of its direct applicability to describing astronomical black hole candidates. In counterpoint, purely mathematically, there is an old classical result of differential geometry, due to Darboux, that all 3-manifolds can have their metrics recast into diagonal form. In the case of the Kerr spacetime the Boyer--Lindquist coordinates provide an explicit example of a diagonal spatial 3-metric. Unfortunately, as we demonstrate herein, Darboux diagonalization of the spatial 3-slices of the Kerr spacetime  is incompatible with simultaneously putting the Kerr metric into unit-lapse form while retaining manifest axial symmetry. 
This no-go theorem is somewhat reminiscent of the no-go theorem to the effect that the spatial 3-slices of the Kerr spacetime  cannot be chosen to be  conformally flat.

\bigskip
\noindent
{\sc Date:} Thursday 3 September 2020; \LaTeX-ed \today

Doran coordinates; ADM decomposition; unit lapse; Darboux diagonalization; \\
3-metric.

\bigskip
\noindent{\sc PhySH:} 
Gravitation
}

%========================================================
\maketitle
%========================================================
\def\tr{{\mathrm{tr}}}
\def\diag{{\mathrm{diag}}}
\def\cof{{\mathrm{cof}}}
\def\pdet{{\mathrm{pdet}}}
\parindent0pt
\parskip7pt
\def\BLrain{{\hbox{\tiny BL-rain}}}
\def\EFrain{{\hbox{\tiny EF-rain}}}
\def\Doran{{\hbox{\tiny Doran}}}

%=====================================================
\section{Introduction}
%=====================================================
The Kerr spacetime~\cite{Kerr,Kerr-Texas,kerr-intro,kerr-book,kerr-book-2} is perhaps the single most astrophysically important and pre-eminent of the currently known exact solutions to the Einstein field equations. 
(For general background discussion see textbooks such as~\cite{Adler-Bazin-Schiffer,D'Inverno,Hartle,Carroll,wald,weinberg, Hobson, MTW}.)
In view of this central physical importance of the Kerr spacetime, it is well worth developing as many distinct viewpoints~\cite{river}, alternative presentations~\cite{doran,natario,rain,Gordon-form,ansatz}, and tractable approximations~\cite{Lense-Thirring,Pfister,4-of-us} to the Kerr spacetime as possible.
We have recently considered various unit-lapse ``rain'' representations of Kerr~\cite{rain}, and now wish to address the interplay between unit-lapse and possible diagonalization of the spatial 3-metric.

It is a classic mathematical result, typically attributed to Darboux~\cite{Darboux:1898}, that under mild conditions 3-manifolds can have their metrics recast in diagonal form~\cite{Darboux:1925,Darboux:1984,Darboux:1991,Darboux:1992,Darboux:2013,Grant:2009}. The question we wish to raise is whether Darboux diagonalization of the spatial slices of the Kerr spacetime is compatible with simultaneously maintaining the unit lapse condition and axial symmetry.
When starting the calculation there were both reasons for optimism and reasons for caution. 
On the one hand:
\begin{itemize}
\item 
Diagonalization of the 3-metric only imposes 3 coordinate conditions on the 4-metric, while the unit lapse condition only adds 1 more. Imposing 4 coordinate conditions in (3+1) spacetime naively seems plausible. 
\item
Boyer--Lindquist coordinates for Kerr explicitly diagonalize the 3-metric, though they are not unit-lapse coordinates.
\item
Many unit-lapse coordinates for Kerr are known to exist~\cite{doran,natario,rain},
including unit-lapse coordinates very closely related to Boyer--Lindquist coordinates~\cite{rain}.
\item
The Lense--Thirring \emph{approximation} to Kerr can explicitly be cast in unit-lapse flat 3-space form~\cite{4-of-us}.
\end{itemize}
On the other hand, the Kerr spacetime sometimes exhibits (perhaps unexpected) obstructions to otherwise plausible conjectures. 
For instance, all static spherically symmetric (and some stationary axisymmetric, \emph{e.g.}, Lense--Thirring) spacetimes can be put into Painlev\'e--Gullstrand~\cite{painleve1,painleve2,gullstrand} form, with a flat spatial 3-metric~\cite{4-of-us,poisson}. But not Kerr. Indeed, the spatial 3-slices of Kerr cannot even be put in conformally flat form~\cite{ValienteKroon:2003,ValienteKroon:2004}.
\enlargethispage{10pt}

Unfortunately in the situation that we are interested in, the Kerr spacetime again is problematic --- 
we shall demonstrate that in the Kerr spacetime the unit-lapse condition is incompatible with diagonalizing the 3-metric while maintaining axial symmetry. This is more than just an effect of incompatible symmetries, (for instance, things do work nicely for Lense--Thirring spacetime), the incompatibility depends on specific properties of the metric coefficients, and so on specific dynamical features arising from solving the Einstein equations.

%-----------------------------------------------------------------------------------------------------
\section{Framework and Setup}
%-----------------------------------------------------------------------------------------------------

All 3 of the fully specific unit-lapse versions of Kerr that we explored in reference~\cite{rain}, (BL-rain, EF-rain, Doran) have spatial 3-metrics of the form:
\begin{equation}
ds^2 = g_{rr}(r,\theta)\,dr^2 +  g_{\theta\theta}(r,\theta)\,d\theta^2 + g_{\phi\phi}(r,\theta)\,d\phi^2
+ 2 g_{r\phi} (r,\theta) \, dr d\phi.
\end{equation}
That is, the pattern of non-zero elements in the spatial 3-metric is:
\begin{equation}
g_{ij} = \left[\begin{array}{ccc} *&0&*\\ 0 &*&0\\ {}*&0&* \end{array}\right]_{ij}.
\end{equation}
We wish to diagonalize this 3-metric, \emph{while preserving unit lapse and manifest axial symmetry}.
To preserve unit lapse, do not touch the time coordinate~\cite{rain}.
To preserve axial symmetry, transformations of the $\phi$ coordinate are limited to
\begin{equation}
\phi=\bar\phi+\Phi(\bar r, \bar\theta); \qquad 
d\phi = d\bar\phi +  \Phi_{\bar r} \;d\bar r + \Phi_{\bar \theta} \;d\bar \theta.
\end{equation}
In principle we could do arbitrary things to $r$ and $\theta$:
\begin{equation}
(r,\theta) \to (\bar r, \bar \theta).
\end{equation}
But let us take the analysis one step at a time.

%-----------------------------------------------------------------------------------------------------
\section{Azimuth-only coordinate transformations}
%-----------------------------------------------------------------------------------------------------
\label{S:azimuth}
%==========================================================

Let us temporarily agree to leave the $r$ and $\theta$ coordinates untouched; only adjust $\phi$.
That is, we consider:
\begin{equation}
r = \bar r; \qquad \theta = \bar\theta; \qquad \phi=\bar\phi+\Phi(\bar r, \bar\theta).
\end{equation}
Thence
\begin{equation}
dr = d\bar r;  \qquad
d\theta = d\bar \theta; \qquad
d\phi = d\bar\phi +  \Phi_{\bar r} \,d\bar r + \Phi_{\bar \theta} \,d\bar \theta.
\end{equation}
So for the 3-metric
\begin{eqnarray}
ds^2 &=& g_{rr}(\bar r, \bar\theta) \,d\bar r^2
+ g_{\theta\theta}(\bar r, \bar\theta)\,d\bar \theta^2
\nonumber\\
&+&  g_{\phi\phi}(\bar r, \bar\theta)\;
 (d\bar\phi + \Phi_{\bar r} \,d\bar r + \Phi_{\bar \theta} \,d\bar \theta)^2
 \nonumber\\
&+&  2 g_{r\phi}(\bar r, \bar\theta)\;
d\bar r
(d\bar\phi + \Phi_{\bar r} \,d\bar r + \Phi_{\bar \theta} \,d\bar \theta).
\end{eqnarray}

Now just pick off the 3 off-diagonal components:
\begin{eqnarray}
g_{\bar r \bar\phi} &=& g_{r\phi}(\bar r,\bar \theta)+ g_{\phi\phi}(\bar r,\bar \theta)\,\Phi_{\bar r};
\\
g_{\bar \theta \bar\phi} &=&  g_{\phi\phi}(\bar r,\bar \theta)\,\Phi_{\bar \theta};
\\
g_{\bar r\bar \theta} &=&
 \Phi_{\bar\theta}\left( g_{\phi\phi}(\bar r,\bar \theta)\,\Phi_{\bar r} + g_{r\phi}(\bar r,\bar \theta) \right).
\end{eqnarray}
We want all 3 of $g_{\bar r \bar\phi} $, $g_{\bar \theta \bar\phi}$, and $g_{\bar r\bar \theta}$ to vanish.
That is, we demand that the 3 PDEs below must be satisfied:
\begin{eqnarray}
E1:&&\qquad\qquad g_{r\phi}(\bar r,\bar \theta)+ g_{\phi\phi}(\bar r,\bar \theta)\,\Phi_{\bar r} =0;
\\
E2:&&\qquad\qquad
g_{\phi\phi}(\bar r,\bar \theta)\,\Phi_{\bar \theta}=0;
\\
E3:&&\qquad\qquad
\Phi_{\bar\theta}\left( g_{\phi\phi}(\bar r,\bar \theta)\,\Phi_{\bar r} + g_{r\phi}(\bar r,\bar \theta) \right)= 0.
 \end{eqnarray}
But, since $g_{\phi\phi}\neq 0$, $E2$ implies $\Phi_{\bar \theta}=0$, which then automatically satisfies $E3$. 

But this also implies $\Phi(\bar r,\bar\theta)\to \Phi(\bar r)$, and $E1$ becomes:
\begin{equation}
{d\Phi(\bar r)\over d\bar r} =
- \; {g_{r\phi}(\bar r,\bar\theta)\over  g_{\phi\phi}(\bar r,\bar \theta)}
\end{equation}
This equation is consistent iff
\begin{equation}
\partial_{\bar\theta} \left(  {g_{r\phi}(\bar r,\bar\theta)\over  g_{\phi\phi}(\bar r,\bar \theta)} \right) = 0.
\end{equation}
But inspection of the BL-rain, EF-rain, or Doran unit-lapse version of Kerr shows that this consistency condition is \emph{not} satisfied. (See the appendix for more details.)
Specifically, (recalling that in this section $\bar r=r$ and $\bar\theta=\theta$), we would want to have
\begin{equation}
\partial_{\theta} \left(  {g_{r\phi}(r,\theta)\over  g_{\phi\phi}(r,\theta)} \right) = 0.
\end{equation}
However
\begin{eqnarray}
\left(  {g_{r\phi}(r,\theta)\over  g_{\phi\phi}(r,\theta)} \right)_{\BLrain} 
&=&{2mar\sqrt{2mr(r^2+a^2)}\over\Delta}\; {1\over\rho^2\Sigma} ;
\\
\left(  {g_{r\phi}(r,\theta)\over  g_{\phi\phi}(r,\theta)} \right)_{\EFrain} 
&=& -  {a \over\Sigma} \left( 1 +{2mr/\rho^2} + \sqrt{2mr/(r^2+a^2)}
\over 1 + \sqrt{2mr/( r^2+a^2)}\right);
\\
\left(  {g_{r\phi}(r,\theta)\over  g_{\phi\phi}(r,\theta)} \right)_{\Doran} 
&=&  -{a\over\Sigma} \sqrt{2mr\over a^2+r^2}.
\end{eqnarray}
\leftline{Here $\rho^2=r^2+a^2\cos^2\theta$, and $\Delta=r^2+a^2-2mr$, while $\Sigma = r^2+a^2 + {2mra^2\over\rho^2} \sin^2\theta$. }
It is the explicit presence of $\theta$ inside $\Sigma$ that is the obstruction to satisfying the consistency condition. 
So no azimuth-only coordinate transformation is capable of diagonalizing the spatial 3-metric.
(We have been rather slow and careful with this calculation to make the general pattern clear.) 

%-----------------------------------------------------------------------------------------------------
\section{Polar-only $(r,\theta)$ coordinate transformations}
%-----------------------------------------------------------------------------------------------------
\enlargethispage{20pt}

Now let us leave the azimuthal coordinate intact, and only transform the  $(r,\theta)$ plane.
That is, we consider:
\begin{equation}
r = G(\bar r, \bar\theta); \qquad \theta = H(\bar r, \bar\theta); \qquad \phi=\bar\phi.
\end{equation}
Thence
\begin{equation}
dr = G_{\bar r} \;d\bar r + G_{\bar \theta} \;d\bar \theta;  \qquad
d\theta = H_{\bar r} \; d\bar r + H_{\bar \theta}\;  d\bar \theta; \qquad
d\phi = d\bar\phi.
\end{equation}
So for the 3-metric
\begin{eqnarray}
ds^2 &=& g_{rr}(G(\bar r, \bar\theta), H(\bar r, \bar\theta))\,
 (G_{\bar r} \,d\bar r + G_{\bar \theta} \,d\bar \theta)^2 
\nonumber\\
&+&  g_{\theta\theta}(G(\bar r, \bar\theta), H(\bar r, \bar\theta))\,
 (H_{\bar r} \,d\bar r + H_{\bar \theta} \,d\bar \theta)^2
\nonumber\\
&+&  g_{\phi\phi}(G(\bar r, \bar\theta), H(\bar r, \bar\theta))\,
 d\bar\phi^2
 \nonumber\\
&+&  2 g_{r\phi}(G(\bar r, \bar\theta), H(\bar r, \bar\theta))\,
(G_{\bar r} \,d\bar r + G_{\bar \theta} \,d\bar \theta) \,d\bar\phi .
\end{eqnarray}

Now to simplify the notation write $g_{ij}(G(\bar r, \bar\theta), H(\bar r, \bar\theta)) \to g_{ij}(\bar r,\bar \theta)$ and then just pick off the 3 off-diagonal components:
\begin{eqnarray}
g_{\bar r \bar\phi} &=&g_{r\phi}(\bar r,\bar \theta)\,G_{\bar r};
\\
g_{\bar \theta \bar\phi} &=&g_{r\phi}(\bar r,\bar \theta) \,G_{\bar \theta};
\\
g_{\bar r\bar \theta} &=&
g_{rr}(\bar r,\bar \theta) \,G_{\bar\theta}  G_{\bar r}
+ g_{\theta\theta}(\bar r,\bar \theta) \, H_{\bar r} H_{\bar\theta}.
\end{eqnarray}

Now we want all 3 of $g_{\bar r \bar\phi} $, $g_{\bar \theta \bar\phi}$, and $g_{\bar r\bar \theta}$ to vanish.

That is we demand that the 3 PDEs below must be satisfied:
\begin{eqnarray}
E1: &&\qquad\qquad
g_{r\phi}(\bar r,\bar \theta)\,G_{\bar r}=0;
\\
E2:&&\qquad\qquad
g_{r\phi}(\bar r,\bar \theta) \,G_{\bar \theta}=0;
\\
E3:&&\qquad\qquad
g_{rr}(\bar r,\bar \theta)\, G_{\bar\theta} G_{\bar r}
+ g_{\theta\theta}(\bar r,\bar \theta) \,H_{\bar r} H_{\bar\theta} = 0.
\end{eqnarray}
But, since we know $g_{r\phi}\neq 0$, then $E1$ implies $G_{\bar r}=0$, and 
$E2$ implies  $G_{\bar \theta}=0$.
But then $G(\bar r,\bar\theta)$ is a constant; 
so it is not a good coordinate --- we have an inconsistency.
(We do not even need to look at equation $E3$.)
So we cannot diagonalize the spatial 3-metric using only polar $(r,\theta)$ coordinate transformations.

%-----------------------------------------------------------------------------------------------------
\section{Axisymmetry preserving coordinate transformations}
%-----------------------------------------------------------------------------------------------------
 
Now consider the general case --- this is the most general thing one can do \emph{without damaging  the manifest axial symmetry}. 
Consider:
\begin{equation}
r = G(\bar r, \bar\theta); \qquad \theta = H(\bar r, \bar\theta); \qquad \phi=\bar\phi+\Phi(\bar r, \bar\theta).
\end{equation}
Thence
\begin{equation}
dr = G_{\bar r} \;d\bar r + G_{\bar \theta} \;d\bar \theta;  \qquad
d\theta = H_{\bar r} \;d\bar r + H_{\bar \theta} \;d\bar \theta; \qquad
d\phi = d\bar\phi +  \Phi_{\bar r} \;d\bar r + \Phi_{\bar \theta} \;d\bar \theta.
\end{equation}
So for the 3-metric
\begin{eqnarray}
ds^2 &=& g_{rr}(G(\bar r, \bar\theta), H(\bar r, \bar\theta))\;
 (G_{\bar r} \,d\bar r + G_{\bar \theta} \,d\bar \theta)^2 
\nonumber\\
&+&  g_{\theta\theta}(G(\bar r, \bar\theta), H(\bar r, \bar\theta))\;
 (H_{\bar r} \,d\bar r + H_{\bar \theta} \,d\bar \theta)^2
\nonumber\\
&+&  g_{\phi\phi}(G(\bar r, \bar\theta), H(\bar r, \bar\theta))\;
 (d\bar\phi + \Phi_{\bar r} \,d\bar r + \Phi_{\bar \theta}\, d\bar \theta)^2
 \nonumber\\
&+&  2g_{r\phi}(G(\bar r, \bar\theta), H(\bar r, \bar\theta))\;
(G_{\bar r} \,d\bar r + G_{\bar \theta} \,d\bar \theta)\;
(d\bar\phi + \Phi_{\bar r} \,d\bar r + \Phi_{\bar \theta} \,d\bar \theta).
\end{eqnarray}

Now to simplify the notation write $g_{ij}(G(\bar r, \bar\theta), H(\bar r, \bar\theta)) \to g_{ij}(\bar r,\bar \theta)$. That is
\begin{eqnarray}
ds^2 &=& g_{rr}(\bar r, \bar\theta)\;
 (G_{\bar r} \,d\bar r + G_{\bar \theta} \,d\bar \theta)^2 
\nonumber\\
&+&  g_{\theta\theta}(\bar r, \bar\theta)\;
 (H_{\bar r} \,d\bar r + H_{\bar \theta} \,d\bar \theta)^2
\nonumber\\
&+&  g_{\phi\phi}(\bar r, \bar\theta)\;
 (d\bar\phi + \Phi_{\bar r} \,d\bar r + \Phi_{\bar \theta} \,d\bar \theta)^2
 \nonumber\\
&+& 2\, g_{r\phi}(\bar r, \bar\theta)\;
(G_{\bar r} \,d\bar r + G_{\bar \theta} \,d\bar \theta)\;
(d\bar\phi + \Phi_{\bar r} \,d\bar r + \Phi_{\bar \theta} \,d\bar \theta).
\end{eqnarray}
Now just pick off the 3 off-diagonal components:
\begin{eqnarray}
g_{\bar r \bar\phi} &=&g_{r\phi}(\bar r,\bar \theta) \,G_{\bar r} 
+ g_{\phi\phi}(\bar r,\bar \theta)\,\Phi_{\bar r};
\\
g_{\bar \theta \bar\phi} &=&g_{r\phi}(\bar r,\bar \theta) \,G_{\bar \theta} 
+ g_{\phi\phi}(\bar r,\bar \theta)\,\Phi_{\bar \theta};
\\
g_{\bar r\bar \theta} &=&
g_{rr}(\bar r,\bar \theta) \,G_{\bar\theta}G_{\bar r}
+ g_{\theta\theta}(\bar r,\bar \theta) \,H_{\bar r} H_{\bar\theta} 
+ g_{\phi\phi}(\bar r,\bar \theta)\,\Phi_{\bar r}\Phi_{\bar\theta} 
\nonumber\\ &&
+ g_{r\phi}(\bar r,\bar \theta) \left[ 
\Phi_{\bar r} \,G_{\bar \theta} 
+\Phi_{\bar \theta} \,G_{\bar r} \right].
\end{eqnarray}
We want all 3 of $g_{\bar r \bar\phi} $, $g_{\bar \theta \bar\phi}$, and $g_{\bar r\bar \theta}$ to vanish.
That is we demand that the 3 PDEs below must be satisfied:
\begin{eqnarray}
E1:&&\qquad\qquad
  g_{r\phi}(\bar r,\bar \theta)\,G_{\bar r}
+ g_{\phi\phi}(\bar r,\bar \theta)\,\Phi_{\bar r}=0;
\\
E2:&&\qquad\qquad
 g_{r\phi}(\bar r,\bar \theta) \,G_{\bar \theta} 
+ g_{\phi\phi}(\bar r,\bar \theta)\,\Phi_{\bar \theta}=0;
\\
E3:&&\qquad\qquad 
g_{rr}(\bar r,\bar \theta) \,G_{\bar\theta}G_{\bar r}
+ g_{\theta\theta}(\bar r,\bar \theta) \,H_{\bar r} H_{\bar\theta} 
+ g_{\phi\phi}(\bar r,\bar \theta)\,\Phi_{\bar r}\Phi_{\bar\theta} 
\nonumber\\ &&
\qquad\qquad\qquad\qquad
+ g_{r\phi}(\bar r,\bar \theta) \left[ 
\Phi_{\bar r} \,G_{\bar \theta} 
+\Phi_{\bar \theta} \, G_{\bar r} \right]=0.
\end{eqnarray}
These are 3 PDEs for 3 unknown functions. This is not, by itself,  necessarily problematic.
However, consider the specific linear combination 
\begin{equation}
 {1\over g_{\phi\phi}(\bar r,\bar\theta)}\; (G_{\bar\theta}\;E1-G_{\bar r}\;E2).
\end{equation}
Then
\begin{equation}
G_{\bar\theta} \,\Phi_{\bar r}- G_{\bar r}\,\Phi_{\bar\theta} = 0.
\end{equation}
This means that the cross product vanishes:
\begin{equation}
(\Phi_{\bar r}, \Phi_{\bar\theta}) \times (G_{\bar r}, G_{\bar\theta}) = 0.
\end{equation}
This implies
\begin{equation}
(\Phi_{\bar r}, \Phi_{\bar\theta}) \propto (G_{\bar r}, G_{\bar\theta}).
\end{equation}
With general solution
\begin{equation}
\Phi(\bar r, \bar\theta) = W( G(\bar r, \bar\theta)),
\end{equation}
for some arbitrary function $W(G)$. 

Now substitute this back into $E1$ and $E2$:
\begin{equation}
E1':\qquad\qquad 
G_{\bar r}(\bar r, \bar\theta) \left[ g_{r\phi}(\bar r, \bar\theta) 
+ g_{\phi\phi}(\bar r, \bar\theta) \,W'(G(\bar r, \bar\theta)) \right]=0.
\end{equation}
\begin{equation}
E2':\qquad\qquad 
G_{\bar \theta}(\bar r, \bar\theta) \left[ g_{r\phi}(\bar r, \bar\theta) 
+ g_{\phi\phi}(\bar r, \bar\theta) \,W'(G(\bar r, \bar\theta)) \right]=0.
\end{equation}
Now we cannot have \emph{both} $G_{\bar r}=0$ and $G_{\bar \theta}=0$, since that would mean $G$ is a constant, and so not a good coordinate. Therefore we must have
\begin{equation}
\left[ g_{r\phi}(\bar r, \bar\theta) + g_{\phi\phi}(\bar r, \bar\theta) \, W'(G(\bar r, \bar\theta)) \right]=0.
\end{equation}
That is:
\begin{equation}
  W'(G(\bar r, \bar\theta)) = - { g_{r\phi}( \bar r, \bar \theta) \over  g_{\phi\phi}( \bar r, \bar \theta) }.
\end{equation}
But unwrapping this in terms of the \emph{original} coordinate system we started with, this means we are demanding
\begin{equation}
  W'(r) = - { g_{r\phi}( r, \theta) \over  g_{\phi\phi}( r, \theta) }.
\end{equation}
But inspection of the BL-rain, EF-rain, or Doran unit-lapse versions of Kerr shows that this consistency condition is not satisfied. (The right hand side is explicitly $\theta$ dependent.)
We have
\begin{eqnarray}
\left(  {g_{r\phi}(r,\theta)\over  g_{\phi\phi}(r,\theta)} \right)_{\BLrain} 
&=&{2mar\sqrt{2mr(r^2+a^2)}\over\Delta}\; {1\over\rho^2\Sigma} ;
\\
\left(  {g_{r\phi}(r,\theta)\over  g_{\phi\phi}(r,\theta)} \right)_{\EFrain} 
&=& -  {a \over\Sigma} \left( 1 +{2mr/\rho^2} + \sqrt{2mr/(r^2+a^2)}
\over 1 + \sqrt{2mr/( r^2+a^2)}\right);
\\
\left(  {g_{r\phi}(r,\theta)\over  g_{\phi\phi}(r,\theta)} \right)_{\Doran} 
&=&  -{a\over\Sigma} \sqrt{2mr\over a^2+r^2}.
\end{eqnarray}
The $\theta$ dependence is hiding in $\Sigma$.
(See appendix for details on these three metrics. Note the very strong similarities to the azimuth-only argument in section \ref{S:azimuth} above.) So no axisymmetry-preserving coordinate transformation is capable of diagonalizing the spatial metric.

Note that knowing the explicit forms of $g_{r\phi}( r, \theta)$ and   $g_{\phi\phi}( r, \theta)$ in \emph{any one of} the three BL-rain, EF-rain, or Doran unit-lapse versions of Kerr is enough to get to this conclusion. 
Also note that we never had to use equation $E3$.
This completes the argument --- in the unit-lapse Kerr context the Darboux diagonalization argument is incompatible with manifest axisymmetry.

%-----------------------------------------------------------------------------------------------------
\section{Discussion and Conclusions}
%-----------------------------------------------------------------------------------------------------

What have we learned from this discussion? On the one hand, this no-go theorem is a very specific mathematical result specifically for the Kerr spacetime. On the other hand, the various ingredients that go into the discussion have a very much wider realm of applicability. For instance,
 the unit-lapse spacetimes occur quite commonly and very naturally in many specific examples of analogue spacetimes~\cite{unexpected, visser:1997, visser:1998, volovik:1999, stone:2001, visser:2001, fischer:2002, novello:2002, probing, Visser:vortex, LRR, Liberati:2005, Weinfurtner:2005, visser:2010, Visser:2013, Liberati:2018, Schuster:2018}.  In an analogue spacetime context the unit lapse condition physically corresponds to a constant signal propagation speed. (This holds, for example, to a good approximation for sound waves in water.) Various analogue spacetimes can then be invoked to help one develop physical intuition in the current more purely general relativistic context, directly relevant to modelling infall and accretion.

Astrophysically, unit lapse versions of stationary spacetimes, (even if they are not vacuum solutions of the Einstein equations), are extremely useful in at they immediately provide an enormous class of timelike geodesics, the ``rain geodesics''. These correspond to the zero angular momentum observers, ZAMOs, that are dropped from spatial infinity with zero initial velocity and zero angular momentum. These rain geodesics  provide an explicit  and quite tractable probe of the spacetime physics. 

The underlying theme behind our attempt at imposing Darboux diagonalization was to continue the search for improved coordinate systems for the Kerr spacetime. Finding such improved coordinate systems (if possible) is  strategically and tactically important for a better understanding of the technically quite challenging and astrophysically important Kerr spacetime; see  particularly the discussion in reference~\cite{kerr-intro}. See also, for instance, recent attempts at finding a ``Gordon form'' for the Kerr spacetime~\cite{Gordon-form}, and recent attempts at upgrading the ``Newman--Janis trick'' from an ansatz to an algorithm~\cite{ansatz}.

Finally, we should also emphasise that the discussion herein also impacts on and informs the discussion and investigation regarding the potential observational ability to distinguish exact Kerr black holes from various ``black hole mimickers'' that have been suggested in the literature --- see for instance references~\cite{small-dark-heavy, BH-in-GR}.  More recently one could consider references~\cite{phenomenology, viability, geodesicaly-complete, pandora, causal, LISA}, and references~\cite{Simpson:2020, Simpson:2019, Simpson:2018, Boonserm:2018, Simpson:2019-core,  
Berry:2020-core-1, Berry:2020-core-2, Bardeen:1968,Hayward:2005,Frolov:2014,Frolov:2017}.

\newpage
%============================================
\appendix
%============================================
%============================================
\section{Appendix: Some explicit metrics}
%----------------------------------------------------------------------------

In this appendix we list some of the explicit metrics utilized in the text.
%============================================
\subsection{Unit-lapse metrics}
%----------------------------------------------------------------------------
The general form for a unit-lapse metric is~\cite{rain}:
\begin{equation}
ds^2 = - dt^2 + h_{ij} (dx^i - v^i dt) (dx^j-v^j dt).
\end{equation}
Then
\begin{equation}
g_{ab} = \left[ \begin{array}{c|c} -1 + (h^{ij} v_i v_j) & -v_j \\  \hline -v_i & h_{ij} \end{array}\right]_{ab};
\qquad \qquad
g^{ab} = \left[ \begin{array}{c|c} -1   & -v^j \\  \hline -v^i & h^{ij} - v^i v^j\end{array}\right]^{ab}.
\end{equation}
Note $\det(g_{ab})=-\det(h_{ij})$ and $g^{tt}=-1$.

%============================================
\subsection{Lense--Thirring (Painlev\'e--Gullstrand version)}
%----------------------------------------------------------------------------
For the Painlev\'e--Gullstrand version of Lense--Thirring~\cite{4-of-us}:
\begin{eqnarray}
\label{E:LT5}
d s^2 &=& - d t^2 +\left(d r+\sqrt{2m\over r} \; dt\right)^2
+ r^2 \left(d\theta^2+\sin^2\theta\; \left(d\phi - {2J\over r^3} dt\right) ^2\right),
\end{eqnarray}
Then
\begin{equation}
g_{ab} = \left[ \begin{array}{c|ccc}
-1+{2m\over r} + {4 J^2\sin^2\theta\over r^4} &  \sqrt{2m\over r} & 0 & -{2J\sin^2\theta\over r}\\
\hline
\sqrt{2m\over r} & 1 & 0 & 0\\
0& 0 & r^2 & 0\\
-{2J\sin^2\theta\over r} & 0 & 0 & r^2\sin^2\theta\\
\end{array}
\right]_{ab}.
\end{equation}
Here $h_{ij}$ is not just diagonal, it is flat 3-space.
\begin{equation}
g^{ab} = \left[ \begin{array}{c|ccc}
-1&  \sqrt{2m\over r} & 0 & -{2J\over r^3}\\
\hline
\sqrt{2m\over r} & 1-{2m\over r} & 0 & \sqrt{2m\over r} \;{2J\over r^3}\\
0& 0 & {1\over r^2} & 0\\
-{2J\over r^3} &  \sqrt{2m\over r}\; {2J\over r^3} & 0 & {1\over r^2\sin^2\theta} - {4J^2\over r^6}\\
\end{array}
\right]^{ab}.
\end{equation}
Note this is unit lapse, $g^{tt}=-1$.

%============================================
\subsection{Boyer--Lindquist}
%----------------------------------------------------------------------------
The standard form for the Boyer--Lindquist version of Kerr is:
\begin{equation}
(g_{ab})_{BL} =  {\left[\begin{array}{c|cc|c}
%1
- 1+{2mr\over\rho^2} & 0 &
 0 & -{2mar\sin^2\theta\over \rho^2}\\ \hline
 %2
 0&{\rho^2\over \Delta}
 &0&0\\
 %3
 0&0&{\rho^2}&0\\ \hline
 %4
- {2mar\sin^2\theta\over \rho^2}&0&
0& 
\Sigma \;\sin^2\theta
\end{array}\right]}_{ab}.
\end{equation}
Note $h_{ij}$ is diagonal.
\begin{equation}
(g^{ab})_{BL} =  {\left[\begin{array}{c|cc|c}
%1
- 1-{2mr(r^2+a^2)\over\rho^2\Delta} & 0 &
 0 & -{2mar\over \rho^2\Delta}\\ \hline
 %2
 0&{\Delta\over \rho^2}
 &0&0\\
 %3
 0&0&{1\over \rho^2}&0\\ \hline
 %4
- {2mar\over \rho^2\Delta}&0&
0& 
{1-2mr/\rho^2\over \Delta\sin^2\theta}
\end{array}\right]}^{ab}.
\end{equation}
\leftline{Here $\rho^2=r^2+a^2\cos^2\theta$, and $\Delta=r^2+a^2-2mr$, while $\Sigma = r^2+a^2 + {2mra^2\over\rho^2} \sin^2\theta$. }
Note this is \emph{not} unit lapse, $g^{tt}\neq -1$.

%============================================
\subsection{Boyer--Lindquist-rain}
%----------------------------------------------------------------------------
For the Boyer--Lindquist-rain version of the Kerr metric $(g_{ab})_\BLrain$ we have~\cite{rain}:
\begin{equation}
{\left[\begin{array}{c|cc|c}
%1
- 1+{2mr\over\rho^2} & \left(1-{2mr\over\rho^2}\right) {\sqrt{2mr(r^2+a^2)}\over\Delta} &
 0 & -{2mar\sin^2\theta\over \rho^2}\\ \hline
 %2
 \left(1-{2mr\over\rho^2}\right) {\sqrt{2mr(r^2+a^2)}\over\Delta}&
{\rho^2\over \Delta} - \left(1-{2mr\over\rho^2}\right) {2mr(r^2+a^2)\over\Delta^2}
 &0&{2mar\sin^2\theta\over \rho^2}  {\sqrt{2mr(r^2+a^2)}\over\Delta}  \\
 %3
 0&0&{\rho^2}&0\\ \hline
 %4
- {2mar\sin^2\theta\over \rho^2}&{2mar\sin^2\theta\over \rho^2}  {\sqrt{2mr(r^2+a^2)}\over\Delta} &
0& 
\Sigma\sin^2\theta
\end{array}\right]}_{ab}
\end{equation}
Note $h_{ij}$ is not diagonal.

For the inverse metric
\begin{equation}
(g^{ab})_\BLrain =  {\left[\begin{array}{c|cc|c}
%1
- 1 & {\sqrt{2mr(r^2+a^2)}\over \rho^2}  &
 0 & -{2mar\over \rho^2\Delta}\\ \hline
 %2
 {\sqrt{2mr(r^2+a^2)}\over \rho^2} &{\Delta\over \rho^2}
 &0&0\\
 %3
 0&0&{1\over \rho^2}&0\\ \hline
 %4
- {2mar\over \rho^2\Delta}&0&
0& 
{1-2mr/\rho^2\over\Delta\sin^2\theta}
\end{array}\right]}^{ab}.
\end{equation}
Note this is unit lapse, $g^{tt}=-1$. Oddly the spatial part of the inverse metric $g^{ij}$ is diagonal, but this is not the same as saying $h_{ij}$ is diagonal.

%============================================
\subsection{Eddington--Finklestein-rain}
%----------------------------------------------------------------------------
In Eddington--Finkelstein-rain coordinates the covariant metric is given by~\cite{rain}:
\begin{equation}
(g_{ab})_\EFrain =
 {\left[\begin{array}{c|cc|c}
%1
- 1+{2mr\over \rho^2} & g_{tr} &
 0 &- {2mar\over\rho^2}\sin^2\theta \\ \hline
 %2
 g_{tr} &g_{rr}
 &0&g_{r\phi}\\ 
 %3
 0&0&\;\;\rho^2\;\;&0\\ \hline
 %4
 -{2mar\over\rho^2}\sin^2\theta&g_{r\phi}&0& 
\Sigma\,\sin^2\theta
\end{array}\right]}_{ab}
\end{equation}
subject to the relatively messy results that
\begin{equation}
g_{rr} = 1+  {a^2\sin^2\theta(2mr/\rho^2) \over
(r^2+a^2)(1+ \sqrt{2mr/(r^2+a^2)})^2};
\end{equation}
\begin{equation}
g_{tr} ={ {2mr/\rho^2} + \sqrt{2mr/(r^2+a^2)} \over 1+ \sqrt{2mr/(r^2+a^2)}};
\end{equation}
\begin{equation}
g_{r\phi} = -a \sin^2\theta \left( 1 +{2mr/\rho^2} + \sqrt{2mr/(r^2+a^2)}
\over 1 + \sqrt{2mr/( r^2+a^2)}\right).
\end{equation}
Note $h_{ij}$ is not diagonal.
The inverse metric is \emph{much} simpler
\begin{equation}
(g^{ab})_\EFrain =  {\left[\begin{array}{c|cc|c}
%1
- 1&{\sqrt{2mr(r^2+a^2)}\over \rho^2}&
 0 & {\sqrt{2mra^2/(r^2+a^2)}\over\rho^2 (1+\sqrt{2mr/(r^2+a^2)})}\\ \hline
 %2
  {\sqrt{2mr(r^2+a^2)}\over \rho^2}&
 {\Delta\over\rho^2}
 &0& {a\over \rho^2}\\
 %3
 0&0&{1\over \rho^2}&0\\ \hline
 %4
 {\sqrt{2mra^2/(r^2+a^2)}\over\rho^2 (1+\sqrt{2mr/(r^2+a^2)})}& {a\over \rho^2}&
 0& {1\over \rho^2\sin^2\theta}
\end{array}\right]}^{ab}.
\end{equation}
Note this is unit lapse, $g^{tt}=-1$.

%============================================
\subsection{Doran}
%----------------------------------------------------------------------------
The Doran metric is~\cite{doran,rain}
\begin{equation}
(g_{ab})_{Doran} = \left[\begin{array}{c|cc|c}
%1
-1 +{2mr\over \rho^2} & \sqrt{2mr\over a^2+r^2} & 
0 & -{2mar\sin^2\theta\over \rho^2}\\
\hline
%2
\sqrt{2mr\over a^2+r^2} & {\rho^2\over r^2+a^2}& 0 & -a\sqrt{2mr\over a^2+r^2} \sin^2\theta\\
%3
0 &0 & \rho^2& 0\\
\hline
%4
-{2mar\sin^2\theta\over \rho^2} & -a\sqrt{2mr\over a^2+r^2} \sin^2\theta & 
0 & 
\Sigma\, \sin^2\theta
\end{array}\right]_{ab}.
\end{equation}
Note $h_{ij}$ is not diagonal.

For the inverse metric
\begin{equation}
(g^{ab})_\Doran = \left[\begin{array}{c|cc|c}
%1
-1 & {\sqrt{2mr(a^2+r^2)}\over \rho^2}  & 0 & 0\\
\hline
%2
{\sqrt{2mr(a^2+r^2)}\over \rho^2} & {\Delta\over \rho^2}& 
0 & {a\sqrt{2mr\over a^2+r^2} \over \rho^2}\\
%3
0 &0 & {1\over \rho^2} & 0\\
\hline
%4
0 &  {a\sqrt{2mr\over a^2+r^2} \over \rho^2} & 
0 & {1\over (a^2+r^2) \sin^2\theta}
\end{array}\right]^{ab}.
\end{equation}
Note this is unit lapse, $g^{tt}=-1$.

%=====================================================
\section*{Acknowledgements}
%=====================================================

JB was supported by a MSc scholarship funded by the Marsden Fund, 
via a grant administered by the Royal Society of New Zealand.
\\
TB was supported by a Victoria University of Wellington MSc scholarship, 
and was also indirectly supported by the Marsden Fund, 
via a grant administered by the Royal Society of New Zealand.
\\
AS was supported by a Victoria University of Wellington PhD Doctoral Scholarship,
and was also indirectly supported by the Marsden fund, 
via a grant administered by the Royal Society of New Zealand.
\\
MV was directly supported by the Marsden Fund, 
via a grant administered by the Royal Society of New Zealand.

%\newpage
%=====================================================
%===================================================== 

%==================================================================
%==================================================================
\end{document}